\documentclass[10pt,preprint]{aastex}

\def\kms{\relax \ifmmode {\ \rm km s}^{-1}\else \ km\ s$^{-1}$\fi}

\def\Mso{{M$_{\rm \odot}$}}

\def\cm3{${\rm cm}^{-3}~$}

\def\nii{[N~{\sc ii}]}

\def\heii{He~{\sc ii}}
\def\oiii{[O~{\sc iii}]}

\def\ha{H$\alpha~$} 
\def\hb{H$\beta$}
 
\slugcomment{}

\shorttitle{Central Stars of PNe in the LMC}
\shortauthors{Villaver, et al.}
  
\begin{document}
  
\title{The Mass Distribution of the Central Stars of Planetary Nebulae in
  the Large Magellanic Cloud
\footnote{Based on observations made with the NASA/ESA Hubble Space
  Telescope, 
obtained at the Space Telescope Science Institute, which is operated by the 
Association of Universities for Research in Astronomy, Inc., under NASA 
contract NAS 5--26555}}
   
\author{Eva Villaver\altaffilmark{2}}
\affil{Space Telescope Science Institute, 3700 San Martin Drive,
Baltimore, MD 21218, USA; villaver@stsci.edu}
\author{Letizia Stanghellini and Richard A. Shaw}
\affil{National Optical Astronomy Observatory, 950 N. Cherry Av.,
Tucson, AZ  85719, USA; lstanghellini@noao.edu,shaw@noao.edu}

\altaffiltext{2}{Affiliated with the Hubble Space Telescope Space Department
of ESA.} 
 
\begin{abstract}

We present the properties of the central stars from a sample of 54 Planetary
Nebulae (PNe) observed in the Large Magellanic Cloud (LMC) with the Hubble
Space Telescope Imaging Spectrograph (STIS). The Hubble Space Telescope's
spatial resolution allows us to resolve the central star from its nebula 
(and line-of-sight stars) at the distance of the LMC, 
eliminating the dependency on photoionization modeling in the determination of
the stellar flux.  For the PNe in  which the central star is detected we obtain
the stellar luminosities by directly measuring  the stellar fluxes through
broad-band imaging and the stellar temperatures through Zanstra analysis. From
the position of the central stars in the HR diagram with respect to theoretical
evolutionary tracks, we are able to determine reliable core masses for 21
central stars. By including the central star masses determined in this paper to
the 16 obtained previously  using the same technique \citep{Vss:03}, we have
increased the  sample of central star masses in the  LMC to 37, for which we 
find a non-Gaussian mass distribution.  The average central star mass for this
sample is $<m_{CS}, LMC>$~=~0.65$\pm$0.07~\Mso,  slightly higher than the one
reported in the literature for both white dwarfs and the central stars of PNe
in the Galaxy.  If significant, this higher average central star mass in the
LMC can be understood in terms of a metallicity dependency on mass-loss rates
during the Asymptotic Giant Branch, since the LMC has on average half the
metallicity compared to the Galaxy.   Finally, for the 37 objects
analyzed in the LMC, we do not find any significant correlation between  the
mass of the central star and the morphology of the nebula.

\end{abstract}

\keywords{Magellanic Clouds --- planetary nebulae: general --- stars: AGB and
  post-AGB --- stars: evolution --- stars: fundamental parameters}

\section{INTRODUCTION}

The central stars of Planetary Nebulae (PNe) are the result of the  evolution
of stars in the approximate mass range 1--8 \Mso~that ascend the Asymptotic
Giant Branch (AGB) after both hydrogen and helium have been exhausted in the
core. The heavy, and largely unknown, mass-loss rates taking place during the
late AGB evolution \citep{Wil:00} ultimately determine the
initial--to--final mass relation for single stars and therefore the mass
boundary between those stars that will evolve 
off the AGB phase into central stars of PNe and ultimately fade into white
dwarfs, and those that will end their lives as Type {\sc ii} Supernovae. 

After the star leaves the AGB, it first evolves into the PN phase at constant
luminosity towards higher effective temperatures following an evolutionary
track that is mostly dependent on the central star mass. The mass of the
central star alone  determines the energy budget and the timescale of evolution
of the nebular shell \citep{Vmg:02}. An accurate measurement of the central
star properties (mass, effective temperature and  luminosity) is then crucial
to understanding how the formation and evolution of PNe relates to the initial
mass 
of the progenitor star.  Moreover, central stars of PNe are the immediate 
stellar precursors of white dwarfs and in principle the determination of their
masses should provide further constraints to fundamental empirical quantities
such as the initial--final mass relation which rely primarily on the
measurements 
of white dwarfs masses. Although the central stars of PNe can be
up to 10$^4$ times  more luminous than white dwarfs, their mass determination
in our Galaxy is extremely uncertain (i.e. \citealt{Sk:89,Setal:00})
primarily due to the poor knowledge of their 
individual distances. The central star evolutionary tracks in the HR diagram
depend primarily on mass and show
very little variation with the central star 
luminosity. 
In the best case scenario, the horizontal part of the HR diagram, the
dependency of the central star mass ($M_{CS}$) with its luminosity ($L_{CS}$)
goes as, $[M_{CS}/M_{\rm  \odot}]~=~0.5+1.8\times 10^{-5}~[L_{CS}/L_{\rm 
\odot}]$ \citep{Vw:94} which means that in  order to determine accurately the
central 
star mass the distance has to be known to better than 10\% \citep{S:06}.

In principle the problem with the distance uncertainties can be overcome by
observing PNe in the Magellanic Clouds, where the distance is known to an
accuracy of $\sim$10\% \citep{Ben:02}, and the distances to all nebulae within
one  Cloud are known to within a few percent.  However, since PNe
in the Magellanic Clouds cannot be resolved using ground-based observations, 
the central star parameters derived from ground--based studies have usually
relied heavily on photoionization modeling of the nebula
\citep{BL,Hlb:89,Dm:91a,Dm:91b,MED,Md:91b}. As a result, these studies can only
determine lower limits to the stellar luminosities \citep{Vill:06}. The high
spatial resolution capabilities of the Hubble Space Telescope ({\it 
HST}) allowed for the first time to spatially resolve PNe in the
Magellanic Clouds. The first {\it HST} observations of PNe in the Magellanic
Clouds, however, were designed to target the nebulae through narrow-band
filters \citep{Detal:94,Detal:96,Detal:97, Vetal:96,
  Vetal:98} and since they could not detect the central star, they did not
yield a significant improvement in the determination of the central star
parameters compared to results based on ground-based observations.

Our group has acquired and analyzed the first {\it HST} data samples of PNe
in the Magellanic 
Clouds, aimed at determining the stellar continuum through broad-band imaging
\citep{Setal:01,Setal:02}. The first detailed analysis of the central stars 
included a sample of 35 PNe in the Large Magellanic Cloud (LMC)
\citep{Vss:03} and 27 PNe in the Small Magellanic
Cloud (SMC) \citep{Vss:04}. Often the central star is not 
detected above the nebular continuum, either because it has evolved to
luminosities below our detection limits or because the nebular continuum is
brighter than the stellar one. As a consequence the properties of only 14 (out 
of 27 PNe observed in the SMC) and 16 (out of 35 PNe in the LMC) central stars
were determined. Note that only four masses of central stars in the LMC were
known previously from direct measurement of the stellar flux
\citep{Detal:93,Betal:97} (seven more central star masses have been
determined in the LMC since then by \citealt{Hb:04}). From the analysis of our
{\it HST} samples we found an average central star mass of
0.63$\pm$0.1~\Mso~in 
the SMC \citep{Vss:04}, which is similar to the average mass obtained in the
LMC (0.65$\pm$0.07~\Mso) \citep{Vss:03}. However, we 
found that the SMC and LMC central star mass distributions are different, in
that the SMC sample lacks an intermediate-mass stellar population (0.65 to
0.75~\Mso) \citep{Vss:04}. 

In this paper, we extend our previous study of the properties of the LMC
central stars to a sample of 54 additional PNe. We present the analysis of the
central stars of PNe  obtained on a Cycle 10 {\it HST} SNAPSHOT survey of LMC
PNe using broad-band imaging with the Space Telescope Imaging Spectrograph
(STIS).  The broad and narrow-band images of the PNe are presented elsewhere
\citep{Shaw:06}, together with the line intensities and nebular physical
conditions obtained  by using {\it HST} STIS slitless spectroscopy. In \S2 we
describe the observations and in \S3 the photometric calibration, and the
central star temperature and luminosity determinations. Our results are
presented in \S4 and discussed and summarized in \S5.

\section{OBSERVATIONS}

The observations presented in this paper are  from the {\it HST} program \#9077
(P.I. Shaw). The targets were scheduled as SNAPSHOT exposures, which means that
the actual observations were selected on the basis of expediency for the
scheduling system, and on the availability of visits with suitable durations.
The observation log, observing configuration, target selection, acquisition and
a description of the  basic calibration (through flat-fielding) can be found in
\citet{Shaw:06}. The target list included all known LMC PNe that have not
already been observed with {\it HST} and was derived from the catalogs of
\cite{Smp:78}, \cite{San:84}, \cite{Jac:80}, \cite{Mg:92} and \cite{Mor:94}. 

The photometry of the central stars has been obtained from the STIS clear
aperture mode images (50CCD). The 50CCD is an unvigneted aperture with a 
field of view of 52\arcsec$\times$ 52\arcsec~ and a focal plate scale of
0\arcsec.0507 pix$^{-1}$. In this setting no filter is used and the shape of
the bandpass is governed by the detector (which has a sensitivity from 
$\sim$2\,000 to 10\,300 \AA) and by the reflectivity of the optics. The
central 
wavelength of the 50CCD is 5850\AA, and the bandpass is 4410 \AA.  The FWHM
of 
the PSF is close to 2 pixels at 5\,000 \AA~and the 90\%  encircled energy
radius 
is 3 pixels \citep{Letal:01}.  The observations  were made with the CCD
detector using a gain of 1 $e^-$ per analog-to-digital  converter unit.
All the exposures were split into two equal  components to allow cosmic-ray
rejection. 

Table~1 gives in column (1) the object name according to the SMP nomenclature
when available, in columns (2) and (3) the sky coordinates, in column (4) the
nebular diameters measured with respect to the 10\% intensity contour of the
outer most structure in the \oiii~5007~\AA~line, in column (5) the total  
integration time, and in column (6) whether or not the central star was
detected in the images.

\section{ANALYSIS}

\subsection{Stellar Photometry}

The photometric technique has already been described in detail in
\cite{Vss:03}. In summary, we have applied aperture photometry using the
IRAF\footnote{IRAF is distributed by the National Optical Astronomy 
Observatory, which is operated by the Association of Universities for Research
in Astronomy, Inc., under cooperative agreement with the National Science
Foundation.} PHOT task. Given the spatially resolved nature of the nebula we
measure the flux within a circular aperture with a radius of 2 pixels centered
on the star, and subtract the  nebular emission by estimating the median
nebular flux in an annulus with a width of 1 or 2 pixels adjacent to the
stellar aperture. The nebular emission within this aperture can be highly
inhomogeneous and strong variations from the median of the nebular flux are
reflected in the large standard deviation. These uncertainties are 
propagated into the errors of the measured magnitudes.  The internal
consistency 
of the procedure, which is accurate at the  1\%  level, has been tested in
previous papers by performing aperture photometry on nebular subtracted images
built by co-adding 2D monochromatic images taken from the STIS spectroscopy. 

The instrumental magnitudes are given in the STMAG\footnote{The STMAG is the
Space Telescope magnitude system, based on a spectrum  with constant flux per
unit wavelength.} system by using the zero-point calibration and aperture
corrections presented in  \cite{Betal:02}. The STIS charge transfer efficiency
(CTE) correction  was applied in this version of the pipeline (see
\citealt{Shaw:06}) but it has been shown only to have an effect less than 0.01
mags \citep{Retal:00} for stars in the center  of the field. 

The stellar extinction correction has been estimated from the nebular Balmer 
decrement, {\it c}, and the relation $c~=~1.41~E_{B-V}$ \citep{Sea:79}, where
E$_{B-V}$ is the color excess.  Ultimately this approach relies on  the lack of
strong spatial variations in the extinction within the nebula caused by
internal absorption by dust. Significant spatial variations in the \hb/\ha
ratio within the nebulae have 
not been found for any of the heavily reddened objects in our sample,
confirming the validity of this approach. The extinction constants have been
taken from \citep{Shaw:06} except for J~5, MG~16 and MO~33 where the extinction
could not be determined because the intensity of the \ha line was blended with
the \nii~6548, 6583 \AA~line emission in the slitless spectra. The extinction
constants for J~5, MG~16, and MO~33 are not available in the literature and
therefore we have assumed zero extinction for these central stars. Note that 
the mean extinction for the bright Magellanic Cloud PNe observed with the {\it
  HST} is $c~=~$0.19 \citep{S:06} so  
adopting a low extinction for this purpose is not unreasonable.  

In the wavelength range under consideration, the LMC and Galactic  extinction
laws are  very similar \citep{How:83}.  Thus,  we used the  interstellar
extinction law of \cite{Sm:79}  and assumed that R$_{V}~=~3.1$.  The extinction
in magnitudes (A$_{V}$) is then A$_{V}~=~2.2 c$.

The transformation from the instrumental STMAG magnitude to the standard V-band
magnitude in the photometric Johnson-Cousins UBVI system  has been derived
using synthetic photometry with IRAF/STSDAS SYNPHOT by using a blackbody
spectrum to represent the Spectral Energy Distribution (SED) of the central
star. The procedure,  which is  described in detail in \cite{Vss:03}, basically
consists in obtaining a median of the V-50CCD colors for blackbodies between
30\,000 and 300\,000 \,K with the extinction values appropriate for each
source. The 
uncertainty in the transformation has been added in quadrature to the error in
the measured magnitude. 

When the central star is not detected (i.e., no stellar PSF appears above the
nebular   level), we computed a lower limit to the central star magnitude by
measuring the flux   inside a stellar aperture at the geometric center of the
nebula (i.e., the  most likely position of the central star). The lower limits
to the magnitudes for those  central stars that have not been detected in our
sample are presented in Table~2 where  column (1) gives the object name,
columns (2) and (3) give  the lower limit magnitude of the central star in the
STMAG and Johnson-Cousins systems respectively and  column (4) gives the
nebular extinction constant.

In Table~3, we give the results of the photometry for the objects in which the
central star has been detected.  Column (1) gives the PN name, columns (2) and
(3) give the STMAG and $V$ magnitudes respectively as well as the associated
errors.  The uncertainties includes the CCD noise (i.e. photon noise and read
noise),   the systematic error (central star flux, sky) and the errors in the
calibration. The nebular extinction constants used to correct for the stellar
extinction are listed in column (4). We have 6 objects (MG~29, Mo~47, Sa~121,
SMP~45, SMP~49, and SMP~74) for which the central star is detected, but at a
marginal level above the nebula (see footnote in Table~1). The large errors
obtained for these objects directly reflect the
uncertainty in the measurement of the stellar magnitude.

\subsection{Effective Temperatures, Bolometric Corrections and Stellar
  Luminosities} 

As in previous papers we have estimated the temperature of the central star by
using the method originally developed by Zanstra \citep{Zan:31} which is
extensively used in the literature (i.e. \citealt{Hs:66,Kal:83,Svmg:02}).  The
Zanstra temperature  is an estimate of the ionizing flux from a star computed
by comparing the flux of a nebular recombination line   of hydrogen or helium
to the stellar continuum flux in the $V$-band assuming a particular choice for
the stellar spectral energy distribution,  which in our case is a blackbody.  

The data needed for the temperature calculation have been taken from
\cite{Shaw:06}  
(\hb~fluxes, nebular radii and extinction constants). The \heii~4686~\AA~line
fluxes were taken from   \cite{BL}, \cite{MED}, \cite{Mp:98}, and \cite{Ld:06}.
In order to assure the best results we have been very   conservative with the
uncertainties in the fluxes quoted by the references.  We have supplemented the
above with fluxes from our unpublished, ground-based NTT spectra for SMP~57. In
Table~3, column (5) we list the \heii~4686~\AA~line intensity relative to
\hb~=~100, not corrected  for extinction; in column (6) we list the reference
code for the \heii~4686~\AA~fluxes.

The bolometric correction (BC) dependence with stellar effective
temperature was taken from 
\cite{Vgs:96} which was derived for Galactic O-type and early B-type stars. We
use this relation because the dependence of BC with $\log$~{\it g} was found
to be extremely weak. The BCs have been computed by using the \heii~Zanstra
temperature when available, otherwise the H~{\sc i} Zanstra temperature was
used.  Temperatures derived from \heii~Zanstra analysis are  the most reliable,
because most PNe are optically thick to  \heii~ionizing photons.  H~{\sc i}
Zanstra temperatures can be reliable for PNe with sufficient optical depth,
but the problem is in determining which PNe are optically thick to hydrogen
ionizing radiation. 

In order to compute the central star luminosities we adopted a distance to
the LMC  of 50.6 {\rm kpc} and an absolute bolometric magnitude for the Sun
of  M$_{bol,\odot}$= 4.75\,mag \citep{All:76}. Note that the same distance to
the LMC is used by the {\it HST} Key Project in the extragalactic distance
scale 
\citep{Freed:01,Moul:00}. From the three dimensional  structure of the LMC
\citep{Fio:83,Vdm:01}, we have estimated a spread in the distance modulus of
0.03, which we propagate into the error of the absolute magnitudes and
luminosities. We have not taken into account the error in the distance to
the LMC, because it will affect all the objects in the same way.

We give the computed temperatures and luminosities in Table~4.  In  column (1)
we give the PN name; in columns (2) and (3) we give the  effective temperatures
(in units of 10$^3$ {\rm K}) derived from the Zanstra method for the \heii~and
for the H~{\sc i} recombination lines,  respectively.  The visual absolute 
magnitudes derived from the BCs are listed in  column (4). The BCs and their
errors, computed by propagating the errors in the determination of the
effective temperature are given in column (5). Column (6) gives the stellar
luminosity and 
column (7) the morphological classification of the PN. We use the morphological
classification of the nebula given in \cite{Shaw:06}. PNe are classified as
Round, Elliptical,  Bipolar, and Point-symmetric according to their morphology 
in the \oiii~5007~\AA~line. We have excluded the bipolar core
morphological class as a secondary morphology indicator given
the small number of objects with bipolar core morphology and with detected
central stars.

\section{RESULTS}

\subsection{Central Star Direct Detection}

We are able to detect and successfully measure a central star
magnitude in 64\% of all PNe presented in this paper, and in
\cite{Vss:03,Vss:04}. When we are unable to measure the central star
magnitude, the most  common reason is that the nebular continuum is much
stronger than that from the central star.  This is illustrated in Figure~1,
where we plot the stellar $V$ magnitudes vs. the average surface brightness
of the PN in H$\beta$~emission. The nebular continuum is
primarily composed of free-bound emission from H$^0$ and He$^+$ (when present),
although free-free emission can be a significant contributor in the blue part
of the spectrum \citep{Osterbrock}. This emission will for the most part be
spatially coincident with the H bound-bound emission. One would expect that
the stellar magnitudes would be well determined above the noise, and less
well determined or undetected as the stellar continuum becomes lost in 
the nebula.  This is clear from Fig.~1, where the error bars on the
magnitudes increase along the upper envelope of the trend line until all
that can be determined 
is a lower limit.  The transition region between well determined magnitudes and
limits is not perfectly sharp because the nebular continuum is often
spatially non-uniform, which can but often does not necessarily complicate
the photometry. Stars shown as upper  limits are those where the central star
was saturated in the image. Stars with very faint lower limits are those that
were not detectable even in the absence of nebular continuum.

In all, the technique of observing the central stars of PNe with the STIS 50CCD
aperture has been very effective and yields an accurate $V$ magnitude for 
most of the time. In cases where the stars are undetectable, useful limits can
be set on the stellar flux. In addition, the high-resolution image (see
\citealt{Shaw:06}) provides a
clear view of the proximity of field stars from which the viability of
ground-based observations of the central star can be judged. Very often it is
the central stars of bipolar nebulae that are lost in the continuum of the
surrounding nebula, which limits our ability to study the evolution of their
progenitors. In order to improve significantly on our technique for detecting
the central star directly, it will be necessary either to obtain high quality
spectra and model the nebular continuum, or to obtain
high-resolution images of these PNe in the ultraviolet.

As mentioned in the introduction, in the past, most of the central star
properties of PNe in the LMC and SMC have been derived from ground-based
optical 
spectroscopic observations that do not permit a separation of the stellar
continuum from that of the nebula. As a result, the central star luminosity and
temperature have to be computed using modeling techniques that rely completely
on nebular measurements. The strongest  implications of this approach is on the
stellar luminosities that are determined from the H$\beta$~line flux of the
nebula under the assumption that the nebula is  optically thick. The nebular
H$\beta$~line flux is used as a measurement of the recombination and therefore
the ionizing flux from the star. The main problem with this approach of
estimating the stellar luminosity is that the nebulae are sometimes totally
optically thin, or thin in some directions, and therefore the derived stellar
luminosities are lower limits.

\subsection{Stellar Distribution on the $\rm \log\,L-\log\,T$ Plane} 

In Figure~2 we show the distribution in the $\rm \log\,L-\log\,T$ plane of
the central stars analyzed in this paper. In the cases where the flux at
\heii~4686~\AA~was 
not available (MG~16, Mo~7, Mo~33, SMP~43, and Mo~47), we have used the
H~{\sc i}~Zanstra temperature to locate the central stars on the HR
diagram. It is well known that the H~{\sc i}~Zanstra temperature
underestimates the  temperature of the central stars of optically
thin PNe (i.e. \citealt{Kj:89, Gv:00}) but it is accurate for optically
thick objects. We have also used H~{\sc i} Zanstra temperatures for J~39 that
has zero measured flux at \heii~4686~\AA\ and for SMP~67 where the measured
\heii~4686~\AA\ flux is very small and therefore the uncertainty is expected
to be very high.

Figure~3 shows the distribution in the $\rm \log\,L-\log\,T$ plane of the total
sample of central stars observed by us with {\it HST} in the LMC. The central
star parameters obtained in \cite{Vss:03} from {\it HST} GO program 8271
\citep{Setal:01} for 16 central stars have been added to the 21 central stars
shown in Fig.~2. We see an identical distribution of the central star
locations in the HR diagram when we plot the two samples together and in
either sample we do not find any central star with an effective temperature
below 30\,000 \,K. From stellar 
evolution theory a massive  progenitor would evolve very fast along this
region of the HR diagram and  therefore the probability of detection is very
small. From numerical simulations coupled to the central star evolution
\citep{Vgm:02,Vmg:02}, a very low-mass central star, although it evolves more
slowly 
in  the HR diagram, it still needs at least 3\,000 \,yr (after the star leaves
the AGB) for a thin PN shell to become ionized and therefore
observable. Shorter timescales than $\sim$3\,000 \,yr are needed to observe
ionized PNe originating from intermediate-mass progenitors \citep{Vmg:02}. We
believe 
that the coupled evolution between the star and the nebula is reflected in
the location of the central stars in the HR diagram shown in Fig.~3.

The central star masses obtained in this paper have been derived from Fig.~2 by
comparing their location on the HR diagram with the \cite{Vw:94} tracks for
stars with LMC  metallicity (Z~=~0.008). Note that the progenitors of the PNe
we studied have a range of initial metallicities, and using single
metallicities for all the tracks has the potential to introduce additional
uncertainty in the determination of individual masses. We have not plotted or
derived masses 
for the stars  that lie below the evolutionary tracks (MG~14, MG~16, Mo~7, Sa
117 and Mo~47). We did not have \heii~4686~\AA~fluxes (or the \heii~4686~\AA~
measured flux is very small) for MG~14, MG~16, Mo~7 and Mo~47 and therefore
their central star parameters are very uncertain. The central star of Sa~117 is
only  marginally detected above the nebula. We did not estimate the central
star masses for MG~45, Sa~104a, and SMP~88  that have much larger luminosities
than those encompassed by the highest mass evolutionary track.  Given the
compact nature of these three PNe, our photometric technique  fails in
subtracting the true nebular contribution from the stellar aperture. J~5 has
been excluded as well from the following analysis because its magnitude was
measured from saturated data and because most of its Balmer emission appears to
be coming from the central star \citep{Shaw:06}. The core masses and  the
morphological classification of the nebulae are summarized in Table~5.   

The \cite{Vw:94} LMC models are more efficient at producing  He-burning
post-AGB tracks for lower mass progenitors, which they argue is a natural
consequence of the mass-loss behavior during the AGB phase.  The He-burning or
H-burning nature of the post-AGB track depends on  whether the star leaves the
AGB when helium-shell or when hydrogen-shell burning is dominant. However, the
mechanism that controls the departure of the star from the AGB is unknown and
therefore artificially defined in the stellar evolutionary models.  From our
data we cannot constrain the nature of  the track, so when possible we have
estimated the masses from both the H-burning and the He-burning LMC tracks.

Note that the central star temperatures and luminosities (and therefore the
masses) have been computed under the assumption of a non-binary central star,
i.e. assuming that all the measured flux arises from the central star. The
reliability of this assumption has been addressed in \cite{Vss:04} where we
explored the possibility of light from a stellar companion contaminating the
photometric measurements. A significant contribution from a stellar companion
to the measured flux in the STIS/50CCD bandpass can be excluded given the
additional restriction that a known distance imposes on the measured flux. It
is important to note that our analysis does not rule out the possibility of a
binary companion to the central star: we only exclude companions with fluxes in
the STIS bandpass that are comparable to that of the ionizing source.

\subsection{Central Star Mass-PN Morphology Relation}

Given the large amount of  observational evidence that shows fundamental
differences in the physical and chemical properties among PN morphological
classes, it has been suggested that the initial mass of the progenitor star
determines the morphology of the PN. In particular, the N and O chemical
enrichment found in the Galactic bipolar and extremely asymmetric  
morphological classes \citep{Gre:71,Pei:78,Tpp:97} together with their lower
average distance from the Galactic plane \citep{Cs:95, Metal:00, Svmg:02,
ParK:06}, suggests that the bipolar class might evolve from more massive
progenitors and given the initial--final mass relation
massive progenitors imply massive central stars. The correlation between the
central star mass and the PN morphology 
has been explored for Galactic samples by several authors
\citep{Scs:93,Amn:95,Gst:97,Svmg:02} who have found slightly different mass
distributions for the central stars of symmetric and axisymmetric PNe. In the
sample of objects analyzed in this paper, we do not find any correlation
between the mass of the central star and the morphology of the nebula.  The 
lack of correlation still holds if we include the LMC objects presented  in
\cite{Vss:03}. The number of objects with bipolar/quadrupolar morphology and
central star masses measured is small (9 objects), although it still represents
24\% of the sample of 37 objects. The average mass of the central stars of
the 9 bipolar PNe is 0.64$\pm$0.04~\Mso. 

In assessing the lack of correlation found between central star mass and
morphology in this sample, it is important to consider the various selection
effects that may shape the outcome. \cite{Shaw:06} pointed out that the 
selection of objects to be imaged with {\it HST} is, for a variety of reasons,
skewed towards brighter objects, and the completeness of faint PNe  in the LMC 
at the time the {\it HST} targets were selected was poor compared to the SMC
\citep{Jd:02}.  Note that the number of PNe known in the in the central 25
sq. degree region of the LMC, especially at the
faint end, have just tripled \citep{Rp:06}. The incidence of bipolars is
higher for less luminous PNe \citep{Shaw:06}, which suggests that  bipolars
might  be under-represented in our sample.  The detection rate of central stars
in bipolar PNe is comparable to or only slightly less than that for the full
sample.  However, the reasons for failing  to detect central stars seems to be
different among the different morphological classes.  While non-detected
central stars in bipolar PNe are  too faint, non-detected central stars in
round and elliptical PNe   are often buried in bright nebulosity (sometimes
even saturated).  Finally, one should recall that  high-mass central stars
evolve very quickly through the HR diagram, and therefore are most likely to be
found at low luminosities. If bipolar nebulae are produced by high-mass
progenitors, then they may fall below our detection limit more readily than
central 
stars in other types of nebulae. The magnitude of these selection effects is
hard to quantify and might play a role in the average central star
mass determined for the bipolar class.

\subsection{The LMC Central Star Mass Distribution}

In Figure~4 we show the histogram of the mass distribution of the 37 central
stars of PNe obtained by us in the LMC (in gray). We have plotted the central
star masses  
obtained in this paper (21 objects)  and the LMC central star masses
from \cite{Vss:03} (16  objects). In the theoretical PN
luminosity function and other PN applications, the central star mass
distribution is generally assumed to be Gaussian.  A Kolmogorov-Smirnov test
\citep{Ks:67} has been applied to determine the likelihood that the sample
has been drawn from a parent population whose underlying distribution 
is normal. The KS test rules out this possibility at the 87\% confidence level.
A similar result is obtained by applying the Shapiro-Wilk test
\citep{Sw:65}.  In \cite{Vss:03}  using a smaller number of objects,  we
pointed out that the measured central star masses were not normally
distributed. Here we corroborate this result using a larger sample.

The mean and the median of the mass distribution of the 37 central stars of
PNe in the LMC are 0.65$\pm$0.07 and
0.64$\pm$0.06~\Mso~respectively. \cite{Svmg:02} analyzed a sample that 
contained $\sim$200 central stars of PNe in the Galaxy and found an
average mass $<~M_{CS}, GAL~>$ = 0.60$\pm$0.13~\Mso. 
The distribution of white 
dwarf masses in the Galaxy is very  narrow as well with a peak at
$\sim$0.57~\Mso~and a tail extending towards  larger masses
\citep{Bsl:92,Fkb:97,Mna:04}. However, it 
has been noted that the average mass of the white dwarf population should be
used with caution, as it depends on the underlying distribution of masses which
is a function of the temperature range covered by the sample \citep{Fkb:97}.
More recently, \cite{Lbh:05} have analyzed a large sample of 350 white dwarfs
from the Palomar Green Survey and found the $<M_{WD}, GAL>$ =
0.603$\pm$0.134 \Mso~in good agreement with the \cite{Svmg:02}
results. The average mass of the total sample of LMC 
central stars presented in Fig.~4 (37 objects) is slightly higher than the
average mass of both white dwarfs and central stars of PNe in the Galaxy
reported in the literature.

The central star mass depends mainly on the stellar mass during the
Main-Sequence phase (hereafter, the initial mass), and on the mass-loss during
the AGB phase. Mass-loss during the AGB phase has a strong  dependency on
metallicity as it is thought to be driven mainly by dust \citep{Wod:79,Bow:88}.
The dust formation process depends on the chemical composition of the gas: the
lower the metallicity the smaller the amount of dust formed,  and the lower the
efficiency of the momentum transfer to the gas. Thus, low metallicity stars
with dust-driven winds are expected to loose smaller amounts of matter
\citep{Wetal:00} and therefore are expected  to end-up with higher central star
masses. It has been shown that mass-loss during the AGB phase can also
occur in the absence of dust \citep{Wil:00}, but in this case the mass-loss
efficiency is much lower. The metallicity of the LMC is on average half that
of the solar 
mix \citep{Rb:89,Rd:90}. 
Therefore, it is well expected that the efficiency of mass-loss during the AGB
phase will be reduced in the LMC. As a result a higher central star
mass should be the outcome of the evolution for a given initial progenitor. 
If we believe the higher average central star mass of PNe we find in the LMC
we might 
have the first observational evidence from PNe progenitors  for reduced
mass-loss rates  in a lower metallicity environment. The consequences are very
important in terms of galactic chemical enrichment, as a higher fraction of
main sequence stars should reach the Chandrasekhar mass limit in the LMC than
in the  Galaxy \citep{Uetal:99,Detal:99,Getal:00}.

It is important to mention that higher masses are obtained from He-burning
evolutionary tracks than from H-burning tracks. As mentioned in \S4.2 the
nature of the tracks in stellar evolution models depends on the artificially
selected point of departure from the AGB, with different authors using
different criteria. Since we cannot constrain the nature of the track from
observations, we have tried to check the consistency of our results with our
best mass estimates. We have selected from the total sample of objects those
with central stars well detected above the nebular level and  with
He~{\sc ii} Zanstra temperatures. For the following analysis we have excluded 8
central stars  located  in the HR diagram with H~{\sc i} Zanstra temperature
determinations (4 central stars from this paper and 4 from \citealt{Vss:03}).
We have excluded as well the 6 central stars marked in Table~1 that have
large errors in the magnitude measurement. This leaves us with 23 central
star masses out of 
the 37 original total sample. Since H-burning tracks
for the initial LMC composition are not available in the literature for the
smaller central star masses (M$_{MS}<$ 2 \Mso), we have estimated the central
star masses using the galactic post-AGB H-burning tracks by \cite{Vw:94}. 
The tracks on the HR diagram for the H-burning stars calculated for the 
Galactic and LMC  initial abundances are very similar, with a small shift
towards higher luminosities for the LMC composition. We  have estimated the 
shift based on the  tracks that we have in common for the Galactic and LMC 
composition, then applied that shift when estimating the H-burning masses of
the low-mass central stars. The core mass-luminosity relation for H-burning
post-AGB stars \citep{Vw:94} has been used to determine the H-burning masses in
all the cases where the central star lies on  the  horizontal part of the
track. These mass determinations are based on fuel consumption  principles and
have a negligible dependency on metallicity. The re-calculated masses from
H-burning tracks for the central stars in this paper are  listed in Table~5.
For some objects we list two mass determinations, the first one listed is the
best determination available from the LMC He-burning tracks at hand and the
second one is the mass determination from the H-burning tracks as explained
above. The masses from H-burning tracks have also been derived for the
central stars presented in \cite{Vss:03}. 

The hatched histogram in Fig.~4 shows the mass distribution obtained for
the reduced and homogeneously determined sample of central stars described
above. It is worth 
noting that the two samples (total and selected) have similar mass
distributions and average masses (0.65$\pm$0.07~\Mso). The average central
star masses, estimated under the assumption that all the stars are H-burners
is very similar to that determined with a mixed of He and H-burners.

\section{CONCLUSIONS}

We have analyzed a sample of 54 PNe in the LMC and obtained reliable physical
parameters for 21 of the central stars in the sample. This is possible given
that the distance to the LMC is known and that {\it HST}'s spatial
resolution allowed us to separate the star from the nebula. 

We derive an average mass  $<m_{CS}, LMC>$~=~0.65$\pm$0.07~\Mso~for the total
sample of central stars analyzed by us in the LMC (37 objects when we combine
the 21 central star masses presented here to the 16 obtained in
\citealt{Vss:03}). This average mass 
suggests that the average central star mass in the LMC is higher than that in
the Galaxy. Although the uncertainties in the
determination  of Galactic central stars masses hamper our conclusions,  if the
initial mass distribution in the galaxies in the 1-5  \Mso~range were the same,
we would expect to find exactly this effect. That is, higher final masses in
the LMC compared to the Galaxy,  which is  a consequence of the reduced
mass-loss rate expected in a lower metallicity environment. 
Note that this holds given that no evidence of a dependency of the IMF with
metallicity has been found. If correct, we might have the
first direct evidence of a mass-loss rate dependency on metallicity found 
from PN progenitors.

We do not find
any significant relation between the morphology and the central star mass in
the LMC sample, although it 
should be noted that there are very few (9) axisymmetric PNe with detected
central stars in the LMC.

It is worth noting that the evolutionary tracks used here \citep{Vw:94} are
calculated based on a single initial composition, while individual  LMC central
stars may span a range of metallicities (see, i.e. \citealt{Setal:00}). This
affects the mass determination by a very small amount, well within the
observational errors, and would not affect our conclusions on the average mass
and the mass distribution. However,  it is important to note that our {\it HST}
sample of LMC objects might be biased towards bright PNe in the
\oiii~5007~\AA\ line (see \citealt{Shaw:06}). If bright PNe in the
\oiii~5007~\AA~are not only the result of evolution but  they also happen to 
host different kind of progenitors,  then our average central star mass has to
be handled with care.

\acknowledgments 
We would like to thank Pierre Leisy for providing us with the \heii~4686
fluxes prior to publication and Meszaros Szabolcs for his help with the 
compilation of the nebular fluxes.

\clearpage
\begin{deluxetable}{lccccc}
\tabletypesize{\scriptsize}
\tablenum{1}
\tablewidth{0pt}
\tablecaption{OBSERVATIONS}
\tablehead{
\multicolumn{1}{c}{}& 
\multicolumn{1}{c}{R.A} &
 \multicolumn{1}{c}{Decl.} &
\multicolumn{1}{c}{Nebular Diameter} &
\multicolumn{1}{c}{Integration} &
\multicolumn{1}{c}{ }\\
\multicolumn{1}{l}{Name}& 
\multicolumn{1}{c}{J(2000)} &
 \multicolumn{1}{c}{J(2000)} &
\multicolumn{1}{c}{(arcsec)} &
\multicolumn{1}{c}{(s)} &
\multicolumn{1}{c}{Central Star Detection}} 
\startdata
J~5          & 5:11:48.05 &  -69:23:42.2 &  0.97 x 1.48  &     120  &     YES     \\ 
J~25	     & 5:19:54.88 &  -69:31:04.3 &  0.45 X 0.32  &     120  &     NO      \\		
J~27 (SMP-48)& 5:20:09.66 &  -69:53:39.2 &  0.40 X 0.36  &     120  &     NO      \\		
J~33 (Sa-115)& 5:21:18.11 &  -69:43:01.9 &  1.53 x 1.79  &     300  &     YES     \\ 
J~34 (SMP-52)& 5:21:23.75 &  -68:35:34.9 &   0.73	 &     120  &     NO      \\		
J~39 (SMP-63)& 5:25:26.12 &  -68:55:55.4 &  0.63 x 0.57  &     120  &     YES     \\	
MG~04        & 4:52:44.83 &  -70:17:50.6 &  4.3 x 3.3	 &     120  &     NO      \\ 
MG~14        & 5:04:27.67 &  -68:58:12.3 &  1.59 x 1.57  &     120  &     YES      \\ 
MG~16        & 5:06:05.22 &  -64:48:48.9 &  1.28 x 1.63  &     120  &     YES     \\ 
MG~29        & 5:13:42.48 &  -68:15:17.9 &  1.48 x 2.30  &     120  &     YES\tablenotemark{a}\\ 
MG~40        & 5:22:35.36 &  -68:24:26.5 &  0.38 x 0.33  &     120  &     YES     \\ 
MG~45        & 5:26:06.45 &  -63:24:04.4 &  0.31 x 0.23  &     120  &     YES     \\ 
MG~51        & 5:28:34.47 &  -70:33:01.9 &  1.22 x 1.43  &     120  &     YES     \\ 
MG~70        & 5:38:12.37 &  -75:00:21.6 &  0.48 x 0.67  &     120  &     NO       \\ 
MO~7         & 4:49:19.86 &  -64:22:36.8 &  0.72 x 0.93  &     120  &     YES     \\ 
MO~21        & 5:19:04.11 &  -64:44:39.3 &  3.1 x 2.9	 &     120  &     NO       \\  
MO~33        & 5:32:09.75 &  -70:24:43.7 &  2.12 x 1.58  &     300  &     YES     \\ 
MO~36        & 5:38:53.81 &  -69:57:56.0 &  1.14 x 0.97  &     120  &     NO      \\ 
MO~47        & 6:13:03.35 &  -65:55:09.4 &   3.47	 &     120  &     YES\tablenotemark{a}\\ 
Sa~104a      & 4:25:32.15 &  -66:47:18.5 &   \nodata     &     120  &     YES      \\  
Sa~107       & 5:06:43.81 &  -69:15:38.4 &   1.70 x 1.62 &     120  &     YES      \\  
Sa~117       & 5:24:56.82 &  -69:15:31.2 &   1.18 x 1.30 &     300  &     YES     \\ 
Sa~121       & 5:30:26.20 &  -71:13:49.5 &   1.58 x 1.65 &     300  &     YES\tablenotemark{a}\\ 
SMP~3       &  4:42:23.79 &  -66:13:01.0 &   0.26 X 0.23 &     120  &     Unresolved \\ 
SMP~5       &  4:48:08.62 &  -67:26:06.5 &   0.46 x 0.50 &     120  &     YES     \\ 
SMP~6       &  4:47:38.97 &  -72:28:21.6 &   0.67 x 0.56 &     120  &     NO      \\ 
SMP~11      &  4:51:37.69 &  -67:05:16.1 &   0.76 x 0.55 &     300  &     NO      \\ 
SMP~14      &  5:00:21.09 &  -70:58:52.3 &   2.41 x 1.87 &     300  &     NO      \\ 
SMP~29      &  5:08:03.34 &  -68:40:16.8 &   0.51 x 0.47 &     120  &     NO      \\ 
SMP~37      &  5:11:03.12 &  -67:47:57.6 &   0.50 x 0.43 &     120  &     NO      \\ 
SMP~39      &  5:11:42.26 &  -68:34:59.3 &   0.60 X 0.55 &     120  &     NO      \\
SMP~43      &  5:17:02.40 &  -69:07:16.1 &   1.11        &     120  &     YES     \\ 
SMP~45      &  5:19:20.75 &  -66:58:08.4 &   1.66 X 1.62 &     120  &     YES\tablenotemark{a}\\ 
SMP~49      &  5:20:09.46 &  -70:25:38.5 &     1	 &     120  &     YES\tablenotemark{a}\\ 
SMP~51      &  5:20:52.56 &  -70:09:36.6 &    \nodata    &      60  &     Unresolved      \\ 
SMP~57      &  5:23:48.69 &  -69:12:20.9 &   0.93 x 0.90 &     300  &     YES     \\
SMP~61      &  5:24:36.37 &  -73:40:40.0 &   0.56 X 0.54 &     120  &     YES     \\
SMP~62      &  5:24:55.24 &  -71:32:56.4 &   0.59 x 0.41 &     120  &     NO      \\ 
SMP~64      &  5:27:35.69 &  -69:08:56.4 &    \nodata    &     120  &     Unresolved \\
SMP~67      &  5:29:16.01 &  -67:32:48.0 &   0.88 x 0.61 &     120  &     YES     \\ 
SMP~68      &  5:29:02.83 &  -70:19:23.8 &   1.33 x 0.97 &     120  &     YES     \\ 
SMP~69      &  5:29:23.31 &  -67:13:21.9 &   1.84 x 1.43 &     120  &     NO      \\ 
SMP~73      &  5:31:22.27 &  -70:40:45.3 &   0.31 x 0.27 &     120  &     NO      \\ 
SMP~74      &  5:33:29.51 &  -71:52:27.9 &   0.79 X 0.63 &     120  &     YES\tablenotemark{a}\\
SMP~75      &  5:33:47.00 &  -68:36:44.0 &   0.56	 &     120  &     NO      \\ 
SMP~82      &  5:35:57.47 &  -69:58:16.6 &   0.31 x 0.30 &     120  &     YES     \\
SMP~83      &  5:36:20.82 &  -67:18:07.0 &   3.98 x 3.63 &     120  &     YES     \\ 
SMP~84      &  5:36:53.17 &  -71:53:39.8 &   0.57 x 0.48 &     120  &     YES     \\ 
SMP~88      &  5:42:33.27 &  -70:29:24.2 &   0.61 x 0.45 &     120  &     YES     \\ 
SMP~89      &  5:42:36.84 &  -70:09:32.0 &   0.51 x 0.45 &     120  &     NO      \\ 
SMP~91      &  5:45:06.02 &  -68:06:49.1 &   1.89 x 1.40 &     120  &     NO      \\
SMP~92      &  5:47:04.63 &  -69:27:34.5 &   0.62 x 0.54 &     120  &     NO     \\
SMP~98      &  6:17:35.59 &  -73:12:37.3 &   0.41	 &     120  &     NO      \\ 
SMP~101     &  6:23:40.24 &  -69:10:38.9 &   1.03 x 0.82 &     120  &     YES     \\  
\enddata
\tablenotetext{a}{Marginal detection above the nebular level} 
\end{deluxetable}

\begin{deluxetable}{lccc}
\tabletypesize{\scriptsize}
\tablenum{2}
\tablewidth{0pt}
\tablecaption{MAGNITUDE LOWER LIMITS}
\tablehead{
\multicolumn{1}{c}{Name}& 
\multicolumn{1}{c}{STMAG}&
\multicolumn{1}{c}{V} & 
\multicolumn{1}{c}{c} \\
\multicolumn{1}{c}{(1)}& 
\multicolumn{1}{c}{(2)}&
\multicolumn{1}{c}{(3)}&
\multicolumn{1}{c}{(4)}}
\startdata
J~25\tablenotemark{a}&$\ge$17.17 &$\ge$16.71   &0.23 \\    
J~27\tablenotemark{a}&$\ge$17.22 &$\ge$16.62   &0.27 \\   
J~34    &$\ge$20.64 &$\ge$20.15   &0.24    \\   
MG~4    &$\ge$27.81 &$\ge$28.10   &\nodata  \\  
MG~70   &$\ge$23.01 &$\ge$22.66   &0.19    \\   
Mo~21   &$\ge$25.89 &$\ge$26.09   &\nodata \\   
MO~36   &$\ge$25.53 &$\ge$24.91   &0.29    \\   
SMP~3   &$\ge$16.96 &$\ge$17.16   &0.01    \\   
SMP~6   &$\ge$23.04 &$\ge$23.38   &0.69    \\   
SMP~11  &$\ge$21.34 &$\ge$20.66   &0.31    \\   
SMP~14  &$\ge$25.56 &$\ge$25.74   &\nodata \\   
SMP~29  &$\ge$18.13 &$\ge$17.72   &0.21    \\   
SMP~37  &$\ge$19.29 &$\ge$18.80   &0.20    \\   
SMP~39  & \nodata   &\nodata      &\nodata \\   
SMP~51  &$\ge$17.52 &$\ge$15.44   &0.86    \\   
SMP~62\tablenotemark{a}  &$\ge$17.50 &$\ge$17.50   &0.07 \\   
SMP~64\tablenotemark{a}  &$\ge$17.05 &$\ge$14.40   &1.12 \\   
SMP~69  & \nodata   &\nodata      &\nodata  \\  
SMP~73  &$\ge$19.64 &$\ge$19.35   &0.17    \\   
SMP~75\tablenotemark{a}  &$\ge$17.02 &$\ge$16.43   &0.26 \\   
SMP~89  &$\ge$18.60 &$\ge$18.15   &0.31    \\   
SMP~91  &$\ge$27.73 &$\ge$28.04   &\nodata \\   
SMP~92  &$\ge$20.04 &$\ge$19.78   &0.16    \\   
SMP~98  &$\ge$18.35 &$\ge$17.74   &0.28     \\  
\enddata
\tablenotetext{a}{Saturated data}
\end{deluxetable}


\begin{deluxetable}{lcccrc}
\tabletypesize{\scriptsize}
\tablenum{3}
\tablewidth{0pt}
\tablecaption{\scriptsize MAGNITUDES, EXTINCTION, AND \heii~FLUXES}
\tablehead{
\multicolumn{1}{c}{Name}& 
\multicolumn{1}{c}{STMAG} &
\multicolumn{1}{c}{V} & 
\multicolumn{1}{c}{c} & 
\multicolumn{1}{c}{I(\heii)} & 
\multicolumn{1}{c}{Reference}\\
\multicolumn{1}{c}{(1)}& 
\multicolumn{1}{c}{(2)} &
\multicolumn{1}{c}{(3)} & 
\multicolumn{1}{c}{(4)} & 
\multicolumn{1}{c}{(5)} & 
\multicolumn{1}{c}{(6)}} 
\startdata
J~5     &17.28$~\pm~$0.05 & 17.57$~\pm~$0.09\tablenotemark{a,b} &\nodata &  40.0$~\pm~$4.0 & 1 \\
J~33    &20.64$~\pm~$0.01  & 20.93$~\pm~$0.07 &  0.00  &  67.0$~\pm~$6.0 & 1\\
J~39    &18.14$~\pm~$0.10  & 18.11$~\pm~$0.15 &  0.14  &   0.0$~\pm~$0.0 & 2\\
MG~14   &20.62$~\pm~$0.02  & 20.91$~\pm~$0.07 &  0.00  &   1.2$~\pm~$0.6 & 3\\
MG~16   &22.46$~\pm~$0.09  & 22.75$~\pm~$0.14\tablenotemark{a} &\nodata &  \nodata& \nodata\\
MG~29  & 24.58$~\pm~$0.18 &  24.39$~\pm~$0.22 & 0.15&  66.7$~\pm~$1.2&3\\

MG~40   &21.21$~\pm~$0.03  & 20.95$~\pm~$0.07 &  0.17  &   3.5$~\pm~$0.5& 3\\
MG~45\tablenotemark{c}   &17.57$~\pm~$0.03  & 15.68$~\pm~$0.04 &  0.78  &   16.1$~\pm~$1.0& 4\\
MG~51   &19.32$~\pm~$0.02  & 19.61$~\pm~$0.07 &  0.00  &   0.8$~\pm~$0.2& 3\\
MO~7    &20.38$~\pm~$0.02  & 20.63$~\pm~$0.07 &  0.01  &  \nodata& \nodata\\
MO~33   &25.29$~\pm~$0.11  & 25.58$~\pm~$0.16\tablenotemark{a} &\nodata & \nodata& \nodata\\ 
MO~47  & 26.51$~\pm~$0.70 &  26.80$~\pm~$0.76\tablenotemark{a}& \nodata &\nodata&\nodata\\
Sa~104a\tablenotemark{c} &17.30$~\pm~$0.02  & 17.36$~\pm~$0.07 &  0.07  &   57.0$~\pm~$3.2& 4\\
Sa~107  &21.24$~\pm~$0.04  & 20.03$~\pm~$0.06 &  0.51  &   3.4$~\pm~$0.2& 3\\
Sa~117  &23.00$~\pm~$0.07  & 22.63$~\pm~$0.10 &  0.21  &   4.9$~\pm~$0.3& 3\\
Sa~121 & 25.52$~\pm~$0.23 &  25.81$~\pm~$0.29 & 0.00&   2.7$~\pm~$1.2&2\\
SMP~5   &17.64$~\pm~$0.03  & 17.83$~\pm~$0.08 &  0.03  &   2.0$~\pm~$0.1& 4\\
SMP~43  &21.39$~\pm~$0.13  & 21.21$~\pm~$0.17 &  0.15  & \nodata & \nodata \\
SMP~45 & 24.21$~\pm~$0.84 &  23.30$~\pm~$0.87 & 0.40&  18.7$~\pm~$1.5&3\\
SMP~49 & 23.26$~\pm~$0.28 &  23.20$~\pm~$0.32 & 0.11&  29.4$~\pm~$1.2&3\\
SMP~57  &20.05$~\pm~$0.04  & 19.74$~\pm~$0.07 &  0.19  & 7.0$~\pm~$2.0 &5\\
SMP~61  &18.52$~\pm~$0.05  & 18.11$~\pm~$0.08 &  0.22  & 1.0$~\pm~$0.2 &4\\
SMP~67  &18.79$~\pm~$0.05  & 18.76$~\pm~$0.10 &  0.14  & 0.2$~\pm~$0.1& 4\\
SMP~68  &22.52$~\pm~$0.10  & 22.80$~\pm~$0.15 &  0.00  & 95.9$~\pm~$9.0& 4\\
SMP~74 & 20.41$~\pm~$0.53 &  20.41$~\pm~$0.58 & 0.09&  27.3$~\pm~$1.3&3\\
SMP~82  &19.16$~\pm~$0.06  & 18.05$~\pm~$0.08 &  0.47  & 54.3$~\pm~$3.1& 4\\
SMP~83  &19.95$~\pm~$0.04  & 20.24$~\pm~$0.09 &  0.00  & 60.1$~\pm~$2.0& 4\\
SMP~84  &17.90$~\pm~$0.03  & 17.92$~\pm~$0.07 &  0.08  & 0.5$~\pm~$0.2& 4\\
SMP~88\tablenotemark{c}  &18.21$~\pm~$0.03  & 16.81$~\pm~$0.04 &  0.58  & 73.0$~\pm~$1.2& 4\\
SMP~101 &19.56$~\pm~$0.03  & 19.60$~\pm~$0.06 &  0.08  & 72.5$~\pm~$2.3& 4\\
\enddata
\tablecomments{1-$\sigma$ errors are quoted throughout.} 
\tablenotetext{a}{No extinction constant available. The $V$-band mag 
was computed assuming zero extinction.} 
\tablenotetext{b}{Saturated data.}
\tablenotetext{c}{Compact PN the quoted
error does not reflect the true uncertainty, which is dominated by the
nebular contribution to the continuum.}

\tablerefs{(1)\cite{BL}; (2) \citep{MED}; (3) \cite{Mp:98}, (3)
 Leisy \& Dennefeld (2006); (5) Our spectra} 
\end{deluxetable}

\begin{deluxetable}{lccrrrc}
\tabletypesize{\scriptsize}
\tablenum{4}
\tablewidth{0pt}
\tablecaption{\scriptsize CENTRAL STAR PARAMETERS}
\tablehead{
\multicolumn{1}{c}{}& 
\multicolumn{1}{c}{T$_{\rm eff}$(He$_{\sc II}$)} & 
\multicolumn{1}{c}{T$_{\rm eff}$(H)}&
\multicolumn{1}{c}{}&
\multicolumn{1}{c}{}&
\multicolumn{1}{c}{}&
\multicolumn{1}{c}{}
\\
\multicolumn{1}{c}{NAME}& 
\multicolumn{1}{c}{(10$^3${\rm K})} & 
\multicolumn{1}{c}{(10$^3${\rm K})}&
\multicolumn{1}{c}{M$_{\rm V}$}&
\multicolumn{1}{c}{BC}&
\multicolumn{1}{c}{$\log L_*/L_{\sun}$}&
\multicolumn{1}{c}{M}\\
\multicolumn{1}{c}{(1)}& 
\multicolumn{1}{c}{(2)} & 
\multicolumn{1}{c}{(3)}&
\multicolumn{1}{c}{(4)}&
\multicolumn{1}{c}{(5)}&
\multicolumn{1}{c}{(6)}&
\multicolumn{1}{c}{(7)}}
\startdata
 J~5     &   62.2$~\pm~$2.9  &   24.3$~\pm~$2.9  &   -0.95$~\pm~$0.09  &  -5.13$~\pm~$0.14 &   4.33$~\pm~$0.07\tablenotemark{a} &E  \\
 J~33    &   86.7$~\pm~$5.4  &   38.7$~\pm~$5.4  &    2.41$~\pm~$0.07  &  -6.11$~\pm~$0.18 &   3.38$~\pm~$0.08 &E  \\
 J~39    &   \nodata         &   44.1$~\pm~$2.0  &   -0.42$~\pm~$0.15  &  -4.11$~\pm~$0.19 &   3.71$~\pm~$0.10 &E  \\
 MG~14   &   58.3$~\pm~$3.8  &   43.6$~\pm~$3.8  &    2.39$~\pm~$0.08  &  -4.94$~\pm~$0.19 &   2.92$~\pm~$0.08 &R  \\
 MG~16   &   \nodata         &   55.5$~\pm~$4.0  &    4.23$~\pm~$0.14  &  -4.79$~\pm~$0.41 &   2.12$~\pm~$0.17 &B  \\
 MG~29   &  173.0$~\pm~$19.9 &  144.5$~\pm~$19.9 &    5.86$~\pm~$0.22  &  -8.17$~\pm~$0.34 &   2.82$~\pm~$0.16 &B  \\
 MG~40   &   64.0$~\pm~$3.1  &   42.0$~\pm~$3.1  &    2.43$~\pm~$0.07  &  -5.21$~\pm~$0.14 &   3.01$~\pm~$0.06 &E  \\
 MG~45\tablenotemark{c}   &   53.7$~\pm~$2.1  &   21.5$~\pm~$2.1  &   -2.84$~\pm~$0.03  &  -4.69$~\pm~$0.12 &   4.91$~\pm~$0.05 &E  \\
 MG~51   &   46.8$~\pm~$1.8  &   26.8$~\pm~$1.8  &    1.09$~\pm~$0.08  &  -4.28$~\pm~$0.12 &   3.18$~\pm~$0.05 &E  \\
 Mo~7    &   \nodata         &   29.7$~\pm~$1.7  &    2.10$~\pm~$0.07  &  -2.94$~\pm~$0.25 &   2.23$~\pm~$0.10 &E  \\
 Mo~33   &   \nodata         &  179.9$~\pm~$0.0  &    7.06$~\pm~$0.16  &  -8.28$~\pm~$0.59 &   2.39$~\pm~$0.24 &E  \\
 Mo~47   &  \nodata          &  243.8$~\pm~$0.0  &    8.28$~\pm~$0.74  &  -9.19$~\pm~$0.97 &   2.26$~\pm~$0.48 &R  \\
 Sa~104a\tablenotemark{c} &   78.0$~\pm~$2.2  &   33.3$~\pm~$2.2  &   -1.16$~\pm~$0.07  &  -5.80$~\pm~$0.08 &   4.69$~\pm~$0.04 &U  \\
 Sa~107  &   56.5$~\pm~$2.4  &   31.4$~\pm~$2.3  &    1.50$~\pm~$0.06  &  -4.84$~\pm~$0.12 &   3.23$~\pm~$0.05 &B  \\
 Sa~117  &   79.4$~\pm~$4.6  &   67.8$~\pm~$4.6  &    4.04$~\pm~$0.17  &  -5.86$~\pm~$0.17 &   2.62$~\pm~$0.08 &P  \\
 Sa~121  &  103.3$~\pm~$10.7 &  190.1$~\pm~$10.7 &    7.28$~\pm~$0.29  &  -6.64$~\pm~$0.31 &   1.64$~\pm~$0.16 &E  \\
 SMP~5   &   54.7$~\pm~$1.1  &   32.7$~\pm~$1.1  &   -0.68$~\pm~$0.08  &  -4.75$~\pm~$0.06 &   4.07$~\pm~$0.04 &E  \\
 SMP~43  &   \nodata         &   66.1$~\pm~$4.0  &    2.69$~\pm~$0.17  &  -5.31$~\pm~$0.45 &   2.95$~\pm~$0.19 &R  \\
 SMP~45  &  132.1$~\pm~$21.3 &  145.9$~\pm~$21.3 &    4.78$~\pm~$0.87  &  -7.37$~\pm~$0.47 &   2.93$~\pm~$0.39 &B  \\
 SMP~49  &  133.7$~\pm~$13.5 &  123.9$~\pm~$13.5 &    4.68$~\pm~$0.32  &  -7.40$~\pm~$0.30 &   2.99$~\pm~$0.8 &R  \\
 SMP~57  &   62.2$~\pm~$3.3  &   33.6$~\pm~$3.3  &    1.22$~\pm~$0.07  &  -5.13$~\pm~$0.16 &   3.46$~\pm~$0.07 &R  \\
 SMP~61  &   57.5$~\pm~$1.6  &   43.8$~\pm~$1.6  &   -0.41$~\pm~$0.08  &  -4.90$~\pm~$0.08 &   4.02$~\pm~$0.05 &E  \\
 SMP~67  &   49.5$~\pm~$2.0  &   42.5$~\pm~$1.0  &    0.23$~\pm~$0.11  &  -3.99$~\pm~$0.18 &   3.40$~\pm~$0.09 &B  \\
 SMP~68  &  160.7$~\pm~$17.0 &  108.6$~\pm~$17.0 &    4.28$~\pm~$0.15  &  -7.95$~\pm~$0.31 &   3.37$~\pm~$0.13 &E  \\
 SMP~74  &  105.2$~\pm~$8.5  &   74.8$~\pm~$8.4  &    1.89$~\pm~$0.58  &  -6.69$~\pm~$0.24 &   3.82$~\pm~$0.25 &E  \\
 SMP~82  &   66.0$~\pm~$3.2  &   25.2$~\pm~$3.2  &   -0.47$~\pm~$0.08  &  -5.30$~\pm~$0.14 &   4.21$~\pm~$0.07 &E  \\
 SMP~83  &  114.8$~\pm~$4.5  &   68.2$~\pm~$4.5  &    1.72$~\pm~$0.09  &  -6.95$~\pm~$0.12 &   3.99$~\pm~$0.06 &B  \\
 SMP~84  &   51.1$~\pm~$2.0  &   37.0$~\pm~$2.0  &   -0.60$~\pm~$0.07  &  -4.54$~\pm~$0.11 &   3.96$~\pm~$0.05 &E  \\
 SMP~88\tablenotemark{c}  &   63.5$~\pm~$2.9  &   22.6$~\pm~$2.9  &   -1.71$~\pm~$0.04  &  -5.19$~\pm~$0.14 &   4.66$~\pm~$0.06 &E  \\
 SMP~101 &   98.2$~\pm~$3.4  &   47.5$~\pm~$3.4  &    1.08$~\pm~$0.06  & -6.49$~\pm~$0.10 &   4.07$~\pm~$0.05 &E  \\
\enddata
\tablecomments{1-$\sigma$ errors are quoted throughout.} 
\tablenotetext{a}{Derived from saturated data.}
\tablenotetext{b}{Compact PN the quoted
errors do not reflect the true uncertainty, which is dominated by the
nebular contribution to the continuum.}
\end{deluxetable}

\begin{deluxetable}{lccl}
\tabletypesize{\scriptsize}
\tablenum{5}
\tablewidth{0pt}
\tablecaption{STELLAR MASSES}
\tablehead{
\multicolumn{1}{c}{Name}&
\multicolumn{1}{c}{Morphology}& 
\multicolumn{1}{c}{M [\Mso]}&
\multicolumn{1}{c}{Comments}}
\startdata
 J~33  & E   &0.60 &   Interpolation He-burning tracks\\
        &     &0.58 &  Estimated from H-burning tracks\\
 J~39   & E   &0.63\tablenotemark{a} &   He-burning track \\
        &     &0.57 &   Estimated from H-burning tracks \\
 MG~29  & B   &0.67 &   Core mass-luminosity relation  \\
 MG~40  & E   &0.54 &   He-burning tracks\\
 MG~51  & E   &0.54 &   He-burning tracks\\
 Mo~33  & E   &0.82\tablenotemark{a} &   Interpolation H-burning track\\
 Sa~107 & B   &0.57 &   He-burning track \\
        &     &0.55 &   Estimated from H-burning tracks \\
 Sa~121 & E   &0.67 &   He-burning track \\
        &     &0.67 &   H-burning track \\
 SMP~5  & E   &0.71 &   Core mass-luminosity relation \\
 SMP~43 & R   &0.53\tablenotemark{a} &   Extrapolation He-burning tracks \\
 SMP~45 & B   &0.61 &   Interpolation He-burning tracks \\
 SMP~49 & R   &0.61 &   Interpolation He-burning tracks\\
 SMP~57 & R   &0.61 &   Interpolation He-burning tracks\\
        &     &0.58 &   Estimated from H-burning tracks\\
 SMP~61 & E   &0.68 &   Core mass-luminosity relation \\
 SMP~67 & B   &0.57\tablenotemark{a} & Interpolation He-burning tracks \\
 SMP~68& E    &0.65 &   Interpolation He-burning tracks\\
        &     &0.65 &   Estimated from H-burning tracks\\
 SMP~74 & E   &0.69 &   He-burning track \\
        &     &0.62 &   Core mass-luminosity relation \\
 SMP~82 & E   &0.80 &   Core mass-luminosity relation  \\ 
 SMP~83 & B   &0.68 &   Core mass-luminosity relation  \\
 SMP~84 & E   &0.67 &   Core mass-luminosity relation  \\
 SMP~101& E   &0.75 &   Core mass-luminosity relation  \\
\enddata 
\tablenotetext{a}{Derived from Hydrogen Zanstra analysis.} 
\tablecomments{Note that the
  masses from He-burning tracks 
  are slightly larger than when derived from H-burning tracks.}
\end{deluxetable}

\clearpage

\begin{figure}
\plotone{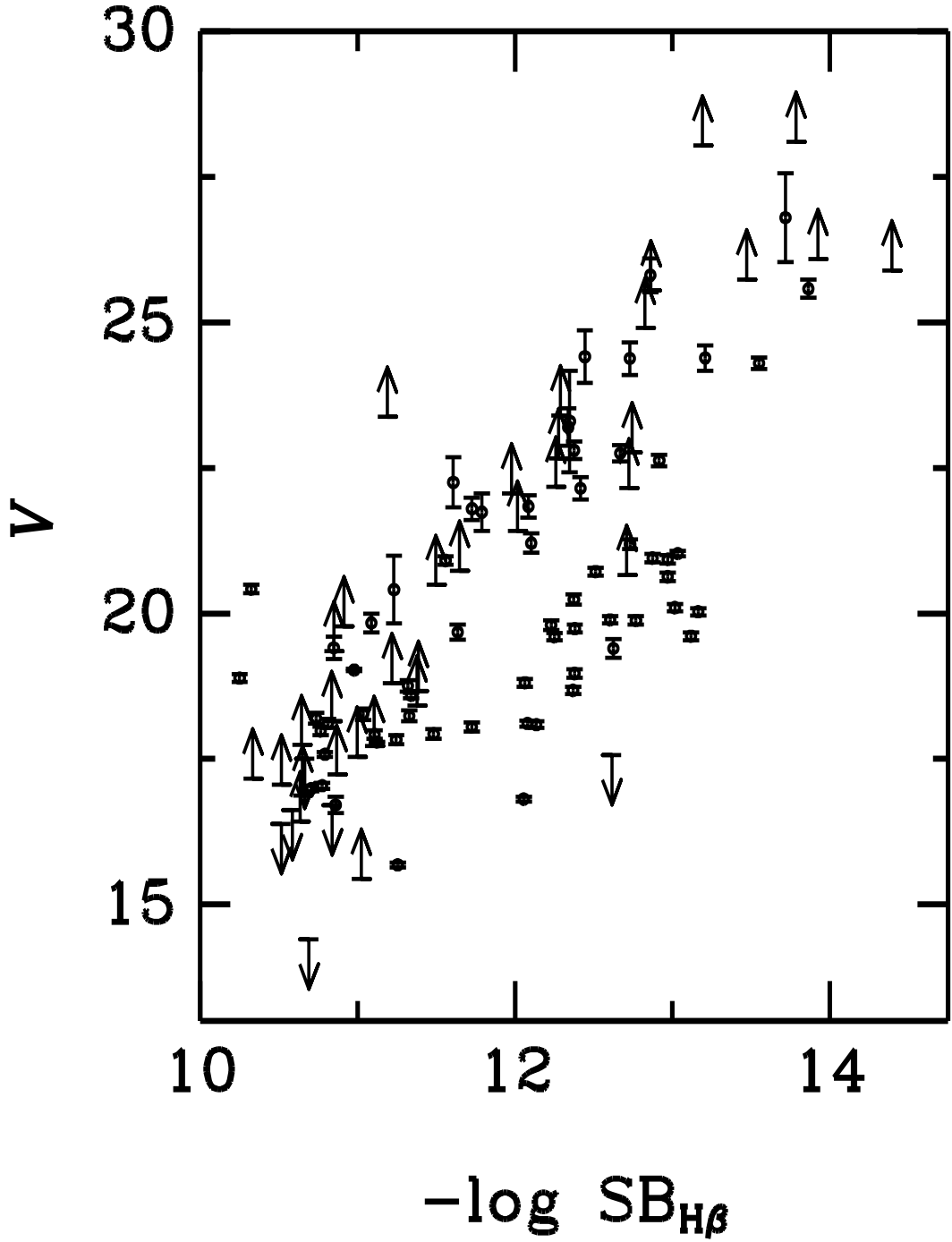}
\caption[ ]{Stellar $V$ magnitudes vs. the average surface brightness of the
PN in the H$\beta$ line. The arrows represents the computed
lower limit to the stellar magnitude when the central star was not detected. 
The upper limits are for saturated objects.
\label{f1.eps}}
\end{figure}

\begin{figure}
\plotone{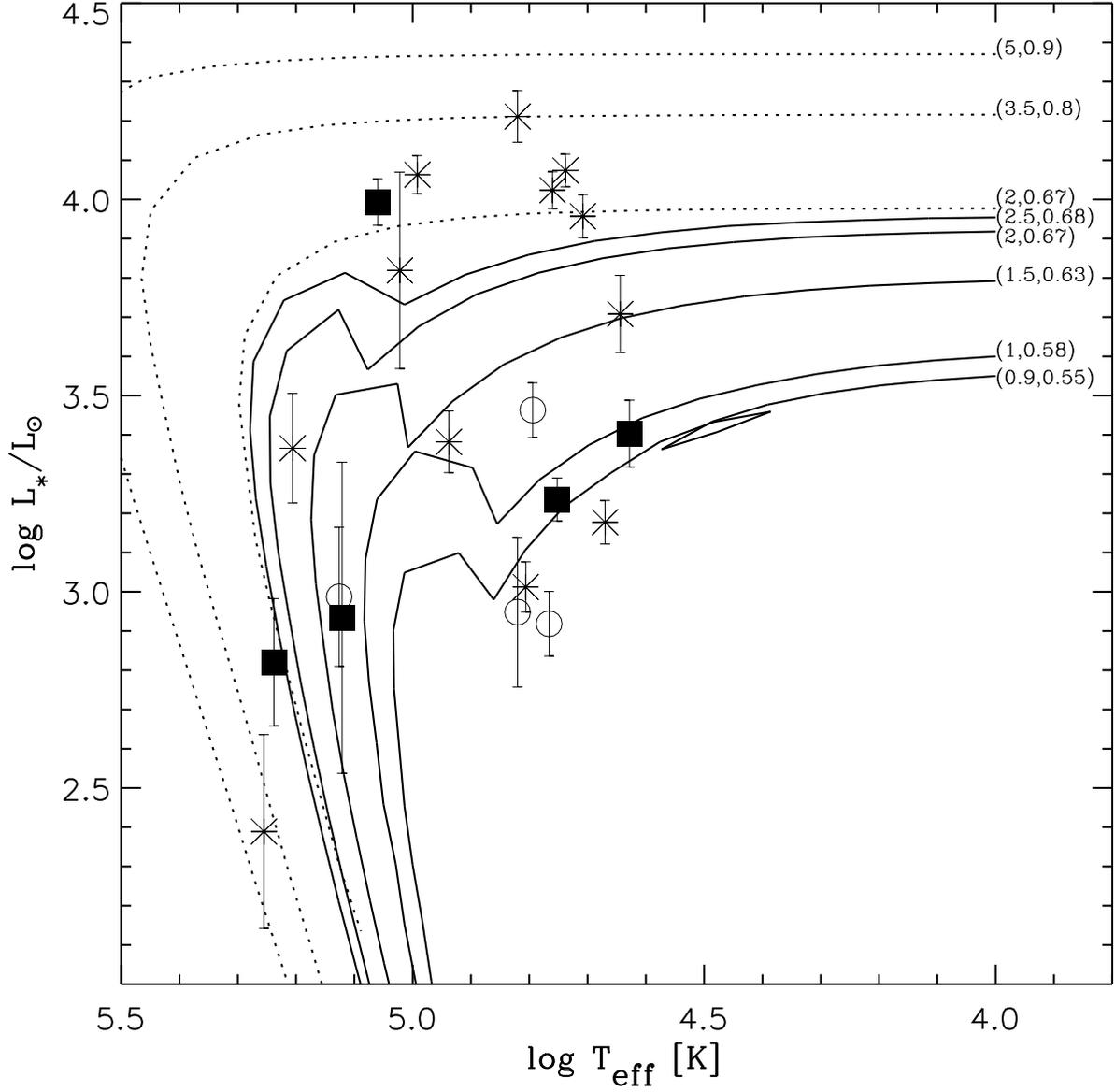}
\caption[ ]{HR diagram for the central stars of PNe analyzed in this
  paper. The symbols represent the 
  morphological types of the hosting  nebulae: round (open
  circles), elliptical (asterisks), bipolar and quadrupolar (squares),
  and point-symmetric (filled circles). Evolutionary tracks are for LMC
  metallicities from 
  Vassiliadis \& Wood (1994), the initial and core masses are marked on
  each track. The solid lines are for He-burners and the dotted lines for
  hydrogen burners. We have only plotted those PN for which we can estimate
  the masses.
\label{f2.eps}}
\end{figure}

\begin{figure}
\plotone{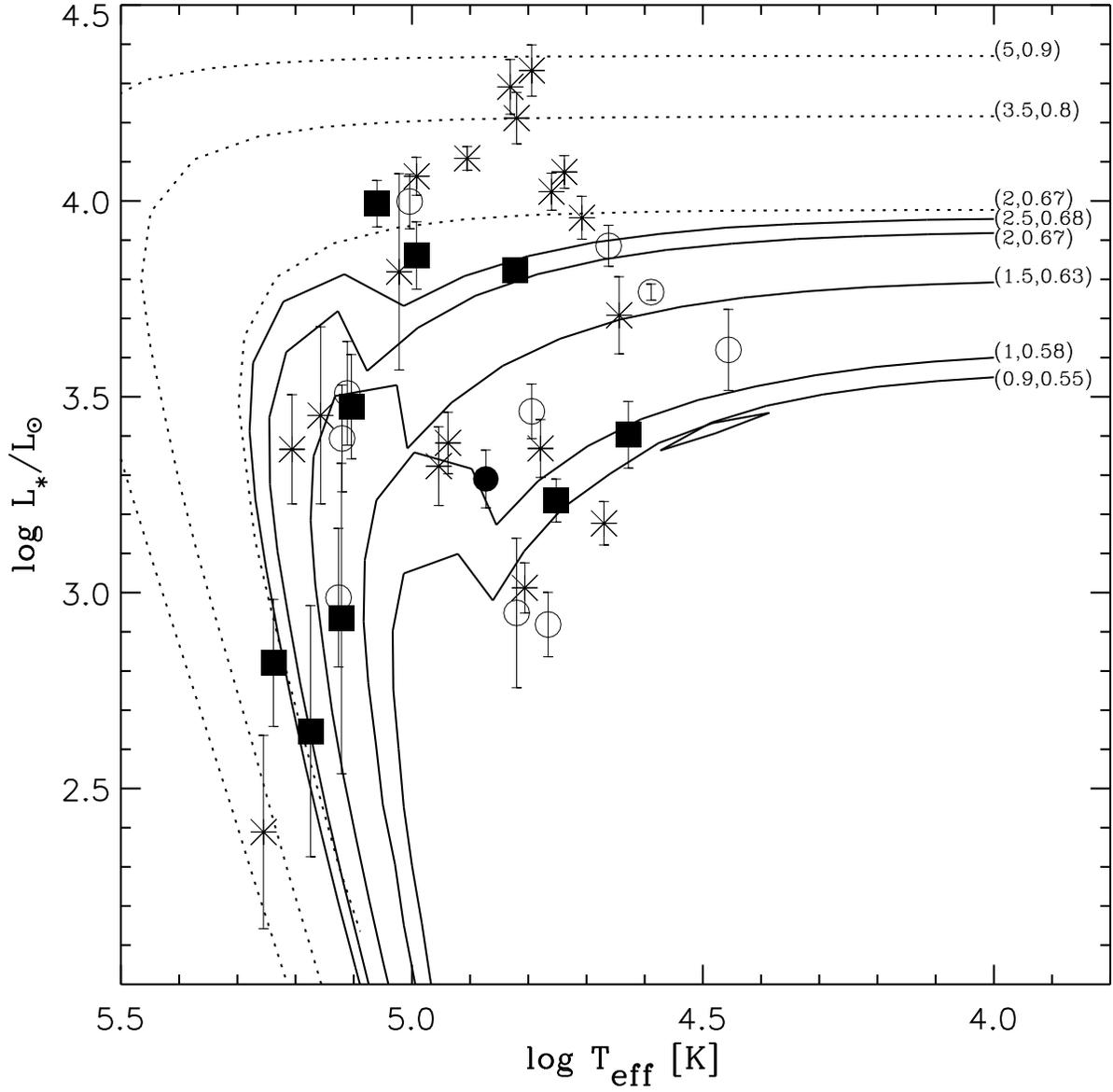}
\caption[ ]{Same as Fig.~1 but including the sample of objects from
  Villaver, Stanghellini \& Shaw(2003). 
\label{f3.eps}}
\end{figure}

\begin{figure}
\plotone{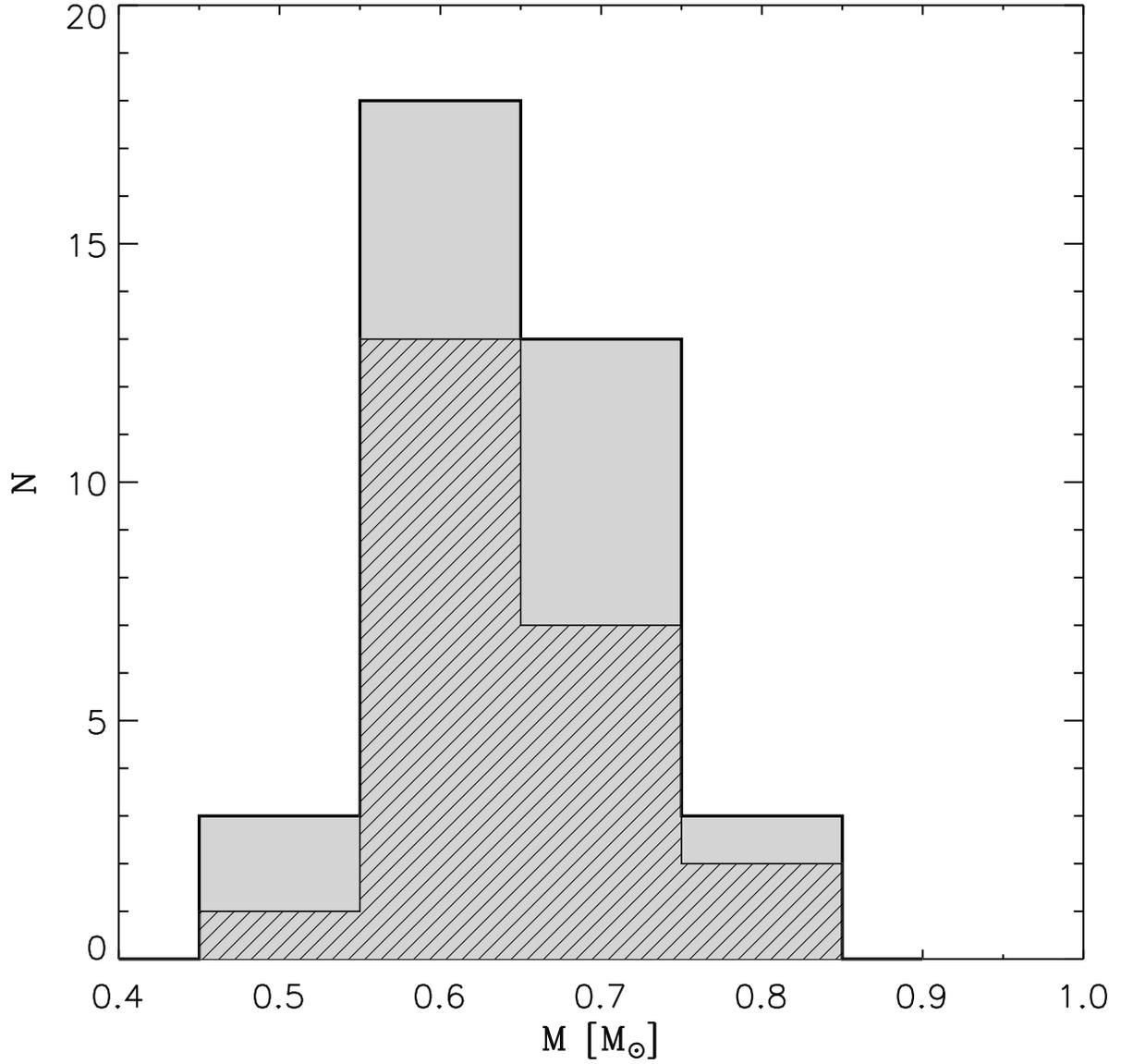}
\caption[ ]{The gray histogram represents the central star mass distribution
  of the sample of objects analyzed by us in the LMC (i.e. masses obtained in
  \cite{Vss:03} and this paper). The hatched histogram represents the mass
  distribution of selected central stars from the whole
  sample that have He~{\sc ii}~Zanstra temperature determinations and masses
  derived 
  from H-burning post-AGB tracks.  

\label{f4.eps}}
\end{figure}

\end{document}